\documentclass[conference]{IEEEtran}
\IEEEoverridecommandlockouts
\usepackage{cite}
\usepackage{amsmath,amssymb,amsfonts}
\usepackage{algorithmic}
\usepackage{graphicx}
\usepackage{textcomp}
\usepackage{todonotes}
\usepackage{xcolor}
\def\BibTeX{{\rm B\kern-.05em{\sc i\kern-.025em b}\kern-.08em
    T\kern-.1667em\lower.7ex\hbox{E}\kern-.125emX}}
\usepackage{graphicx}
\usepackage{textcomp}
\usepackage{bytefield}
\usepackage{rotating}
\usepackage{subcaption}
\usepackage{multirow}
\usepackage{color,soul}
\usepackage{url}
\usepackage{balance}
\usepackage{pifont}
\usepackage[font=small]{caption}
\usepackage{listings}
\usepackage{amsmath}
\usepackage{amssymb}
\usepackage{multirow}
\usepackage{amsthm}
\usepackage{xcolor}
\usepackage{enumitem}
\usepackage{nameref}
    \usepackage{mathtools}

\definecolor{keywordsColor}{rgb}{0.8, 0, 0}    
    
\colorlet{punct}{red!60!black}
\definecolor{background}{HTML}{EEEEEE}
\definecolor{delim}{RGB}{20,105,176}
\colorlet{numb}{magenta!60!black}

\lstdefinelanguage{json}{
    backgroundcolor=\color{white},
}    
    
\begin{document}

\title{TSNZeek: An Open-source Intrusion Detection System for IEEE 802.1 Time-sensitive Networking}

\author{\IEEEauthorblockN{Do\u{g}analp Ergen\c{c}, Robin Schenderlein, Mathias Fischer}
\IEEEauthorblockA{\textit{University of Hamburg}, Germany \\
name.surname@uni-hamburg.de}
}

\maketitle

\begin{abstract}
IEEE 802.1 Time-sensitive Networking~(TSN) standards are envisioned to replace legacy network protocols in critical domains to ensure reliable and deterministic communication over off-the-shelf Ethernet equipment. However, they lack security countermeasures and can even impose new attack vectors that may lead to hazardous consequences. This paper presents the first open-source security monitoring and intrusion detection mechanism, TSNZeek, for IEEE 802.1 TSN protocols. We extend an existing monitoring tool, Zeek, with a new packet parsing grammar to process TSN data traffic and a rule-based attack detection engine for TSN-specific threats. We also discuss various security-related configuration and design aspects for IEEE 802.1 TSN monitoring. Our experiments show that TSNZeek causes only $\sim$5\% CPU overhead on top of Zeek and successfully detects various threats in a real TSN testbed. 
\end{abstract}

\section{Introduction}

% mission-critical system requirements
Modern mission-critical systems are composed of several interconnected components and services that require reliable and time-sensitive communication. 
% presenting IEEE TSN
To satisfy such requirements and reduce the dependency on domain-specific networking technologies, IEEE 802.1 Time-sensitive Networking~(TSN) task group has proposed a set of standards. These standards enable low-latency, fault-tolerant, and deterministic communication on top of standard Ethernet protocols.
% security
However, IEEE 802.1 TSN protocols also induce several security threats across the domains and systems deploying them~\cite{Ergenc2021b, network-anomaly-detection, dos}. Timely detection of such threats is crucial, especially in safety-critical systems, in which a successful attack may lead to hazardous results. 

% the problem
Since the TSN standards are relatively new and prioritize the tight quality of service~(QoS) requirements of critical systems, there is no comprehensive security solution against TSN-specific threats. 
% insufficiency of the existing solution
Although a built-in traffic policing protocol is a part of the standards~\cite{ieee-802.1qci}, it provides limited filtering capabilities to ensure that critical data streams receive sufficient resources. 
% the requirements - monitor and intrusion detection
However, an effective solution requires monitoring time-sensitive streams over TSN protocols and recognizing malicious attempts. 
% our proposal
Accordingly, in this paper, we introduce an open-source network monitoring and intrusion detection system for IEEE 802.1 TSN protocols, \texttt{TSNZeek}. We extended an existing monitoring tool, Zeek (former Bro~\cite{Bro1998}), to analyze the new TSN protocols and detect several TSN-specific attacks. Zeek is a well-established open-source network monitoring solution that is popularly deployed in real systems as well as used in academia for research purposes\footnote{Zeek Project, https://zeek.org}. 
% protocols
We focus on the threats against two TSN protocols, IEEE 802.1CB Frame Replication and Elimination for Reliability~(FRER) and IEEE 802.1Qcc Stream Reservation Protocol~(SRP), since they (i) are the critical protocols addressing communication reliability and configuration, and (ii) have specific network behavior with new packets types and architectural aspects.
% contributions
Our contributions are listed as follows:
\begin{itemize}[leftmargin=*]
\item We implement a new packet parser using a grammar definition language, \textit{spicy}, to process SRP and FRER traffic via Zeek. To the best of our knowledge, this renders \texttt{TSNZeek} the first TSN-aware security monitoring tool.
\item We implement an intrusion detection engine connected to Zeek to recognize several attacks described in our previous paper~\cite{Ergenc2021b}.
\item We test our proposal in a TSN testbed and confirm that it successfully detects various threats with negligible overhead. 
\item We publish \texttt{TSNZeek} open-source together with the traffic and attack generation tools\footnote{The source code is available at \url{https://github.com/UHH-ISS/tsnzeek}}.
\end{itemize}

The remainder of this paper is organized as follows. Section~\ref{sec:background} introduces the TSN protocols that \texttt{TSNZeek} is capable of processing, SRP and FRER. Section~\ref{sec:related} presents the related work. Section~\ref{sec:TSNZeek} describes the design and implementation of \texttt{TSNZeek}. Section~\ref{sec:experiments} gives the experiment setup and evaluation. Lastly, Section~\ref{sec:conclusion} concludes the paper.

\section{Background on IEEE 802.1 TSN} \label{sec:background}

This section describes two TSN protocols: IEEE 802.1Qcc SRP and IEEE 802.1CB FRER.
% why are they important
In comparison to other TSN protocols, SRP and FRER introduce their own interfaces, packet structures, and configuration schemes that impose several security threats. Therefore, we mainly focus on them in this study.

\subsection{IEEE 802.1Qcc Stream Reservation Protocol~(SRP)} \label{sec:srp}

% what is SRP
IEEE 802.1Qcc SRP introduces the resource reservation routines for time-sensitive streams to configure all TSN components in the systems satisfying tight QoS requirements. 
It proposes two main components: (i) a network configuration~(CNC) entity to configure the TSN bridges remotely and (ii) a user configuration~(CUC) entity to discover the endpoints~\cite{ieee-802.1qcc}.
% different configuration scheme
It further offers three configuration schemes utilizing those entities.
\begin{itemize}[leftmargin=*]
% centralized
\item In the \textbf{fully centralized model}, endpoints directly communicate with CUC over a user/network interface~(UNI) and request network resources for TSN streams with certain requirements such as the worst-case latency and inter-arrival times. CNC then configures the bridges according to the requests received by CUC.
% hybrid model
\item In the \textbf{centralized network/distributed user model}, edge TSN bridges, e.g., bridges that endpoints are directly attached to, forward SRP requests to CNC with network-wide visibility. Similar to the fully centralized model, it is responsible for configuring all TSN bridges.
% distributed model
\item In the \textbf{distributed model}, TSN bridges forward SRP requests to each other to handle configurations individually. 
\end{itemize}

% packet structure
SRP imposes a complex packet structure that allows an endpoint to specify various requirements via type-length-value~(TLV) fields and recursive header groups. The respective standard~\cite{ieee-802.1qcc} at Section 35.2 (p.105-134) explains the whole packet structure in detail.

\subsection{IEEE 802.1CB Frame Replication and Elimination for Reliability (FRER)}

IEEE 802.1CB FRER enables redundancy against link failures by sending duplicate TSN flows, which are called member streams~\cite{ieee-802.1cb}. 
% basics
The talker sends member streams through multiple redundant paths configured in advance. 
% sequence number
An incremental sequence number is embedded in FRER frames within the R-TAG header, and the duplicate frames across the member streams have the same sequence number.  
% rejoin
The member streams rejoin at one or more points (e.g., at the listener or an edge bridge) in the network, where duplicate frames are discarded by their sequence number. Finally, the listener receives the original compound stream.

% recovery and timeout
To discard the duplicate frames and obtain the original stream, FRER utilizes various stream recovery functions. These functions consider the sequence number of the most recently received frame to perform frame elimination. For instance, the match recovery function eliminates all the frames with a sequence number smaller than the recently observed one. 
% elimination
The frame elimination helps to drop the duplicate packets received due to stuck senders or misrouting.

\section{Related work} \label{sec:related}

In this section, we briefly present related work that investigates the security threats against IEEE 802.1 TSN protocols and proposes security solutions.

% security of TSN
Regarding security threats, in \cite{Fischer2022}, the authors discuss the security threats in TSN-based industrial control systems. In \cite{Topsakal2022}, they analyze the impact of denial-of-service attacks on TSN protocols. In~\cite{Ergenc2021b}, we listed more than 30 attack vectors against several TSN mechanisms.

% using TSN protocols to prevent attacks, security for TSN
Some of the existing TSN protocols can be utilized against such security threats.
% cbs
For example, in~\cite{dos}, the authors employ IEEE 802.1Qav Credit-based Shaper~(CBS) to prevent denial of service attacks.
% psfp
The authors of~\cite{Meyer2020} and \cite{network-anomaly-detection} combine IEEE 802.1Qci Per-Stream Filtering and Policing~(PSFP) protocol~\cite{ieee-802.1qci} with a centralized controller to enforce ingress policies for packet inter-arrival times and rates, and stream bandwidth. Similarly, in~\cite{luo2021security}, the effectiveness of PSFP is analyzed for the security of TSN-based automotive networks. Lastly, in~\cite{Barton2018}, the authors discuss security policies via PSFP enforced by a centralized policy server. 

There are more practical design and implementation efforts for the monitoring and protection of time-sensitive systems.
% monitoring
In \cite{TSN-insight}, the authors propose a monitoring system for the bridge and link status as well as time-synchronization accuracy, excluding security and intrusion detection aspects. 
% hardware/encryption
\cite{Muguira2020} presents a security module to improve TSN protocols with hardware encryption and authentication. IEEE 802.1AE Media Access Control~(MAC) standard also enables authentication, integrity, and confidentiality in Ethernet-based data traffic~\cite{ieee-802.1ae}.

None of the works above offers security monitoring or an IDS for IEEE 802.1 TSN protocols with specific packet structures, traffic characteristics and requirements, and thus security threats. In contrast, we propose an open-source and extendable IDS solution to address TSN-specific attacks.

\section{TSNZeek: Design and Implementation} \label{sec:TSNZeek}

\begin{figure*}[ht!]
  \centering
  \includegraphics[scale = 0.52]{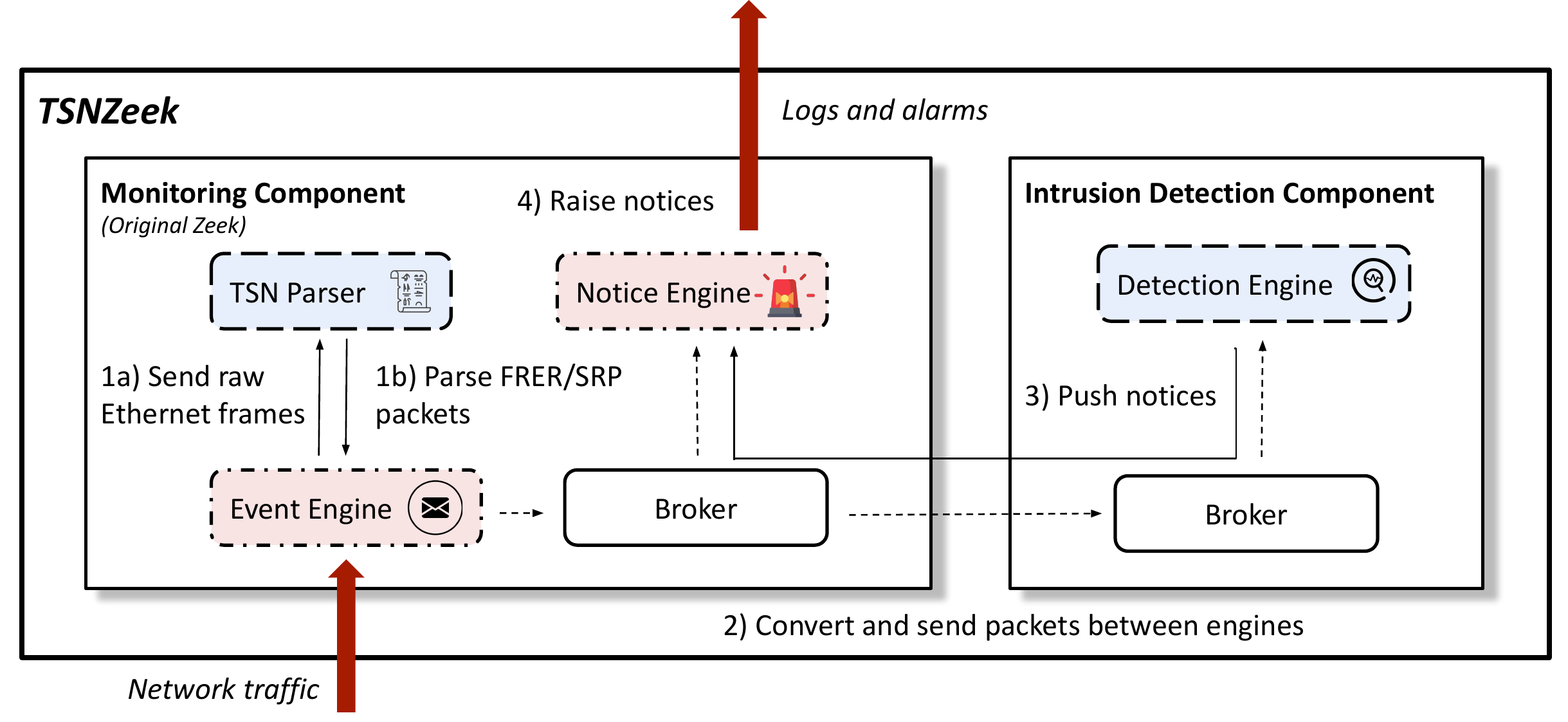}
    \caption{The overview of \texttt{TSNZeek}. The blue/dashed blocks have been implemented from scratch. The red/dotted blocks are existing Zeek modules that we have extended and reconfigured. \vspace*{-0.15in}}
  \label{fig:architecture}
\end{figure*}

\texttt{TSNZeek} consists of monitoring and intrusion detection components shown in Fig.~\ref{fig:architecture}. The monitoring component processes and log the received TSN traffic. The intrusion detection component obtains the processed frames from the monitoring component and implements the attack recognition logic for TSN-specific attacks.

% brief flow
The overall operation of \texttt{TSNZeek} can be described as follows: the \textit{event engine} distinguishes the incoming packets by their EtherType values, which is a standard header type of Ethernet frames. Then, \textit{TSN parser} processes SRP and FRER packets according to the new parsing grammar that we developed. After parsing, the \textit{broker} disseminates those frames to the \textit{notice engine} and \textit{detection engine}. While the former engine logs the traffic and specific events using built-in Zeek functionalities, and the latter implements our attack detection rules. The detection engine then pushes notices back to the notice engine when an attack is recognized. 

In the rest of this section, we elaborate on those modules together with the notices and alerts that \texttt{TSNZeek} can raise. 

\subsection{Monitoring Component}

The monitoring component corresponds to the original Zeek in terms of its working dynamics. It consists of the event engine, the TSN parser, the broker, and the notice engine. The TSN parser (blue, dashed lines in Fig.\ref{fig:architecture}) has been implemented from scratch. The event and notice engines (red, dashed, and dotted lines) are existing Zeek components that we extended further. For the engines, we used Zeek v4.2.0.
 
\subsubsection{Event Engine}
The event engine registers the protocol analyzers during initialization to specify the parsing grammar for respective protocols. Fig.~\ref{fig:zeek-init-1} shows a sample registration process for SRP and FRER. When an SRP or FRER frame is recognized by its Ethernet header (e.g., with type \texttt{0x22EA} and \texttt{0xF1C1}, respectively), the event engine calls the respective packet parsing function implemented in the \textit{TSN parser}.

\begin{figure}[h]
\begin{lstlisting}[language=Python, linerange={5-9}, basicstyle=\footnotesize]
event zeek_init() {
	if (!PacketAnalyzer::try_register_packet_analyzer_by_name("Ethernet", 0xF1C1, "spicy::FRER"))
		print "Cannot register IEEE 802.1CB FRER analyzer";
	if (!PacketAnalyzer::try_register_packet_analyzer_by_name("Ethernet", 0x22EA, "spicy::SRP"))
		print "Cannot register IEEE 802.1Qcc SRP analyzer"; }
\end{lstlisting}
\caption{Registration of packet parsers. } \label{fig:zeek-init-1}
\end{figure} 

The event engine registers further events to trigger logging facilities and inter-module packet dissemination via the broker. Both use the frame content provided by the protocol analyzers, i.e., for the received and parsed SRP and FRER frames.
% referring to the source
In the source code, all event registrations are implemented in the scripts \textit{main.zeek} and \textit{tsn.zeek} via Zeek scripting language.

\subsubsection{TSN Parser} \label{sec:parser}
This module introduces the parsing functions for complex packet structures with many recursive header types. 
% \textit{spicy}
We implemented the parser using \textit{spicy} v1.4.0, which is a grammar generation framework for network protocols and file formats\footnote{Zeek spicy, https://docs.zeek.org/projects/spicy}. 
% standard compliance
We follow the packet definitions in the standards of SRP and FRER so that our parser can process any TSN traffic originated from a standard-compliant talker.

\begin{figure}[t!]
\begin{lstlisting}[language=C, firstline=398, lastline=409, basicstyle=\footnotesize]
function makeTalker(obj: SRP::Talker): TSNZeek::Talker {
	local lendStationInterfaces       = makeEndStationInterfaces(obj.endStationInterfaces);
	local ldataFrameSpecification     = makeDataFrameSpecification(obj.dataFrameSpecification);
	local ltSpecTimeAware             = makeTSpecTimeAware(obj.tSpecTimeAware);
	local luserToNetworkRequirements  = makeUserToNetworkRequirements(obj.userToNetworkRequirements);
	local linterfaceCapabilities      = makeInterfaceCapabilities(obj.interfaceCapabilities);
	return (
		makeStreamID(obj.streamID), makeStreamRank(obj.streamRank),
		lendStationInterfaces, ldataFrameSpecification,
		makeTrafficSpecification(obj.trafficSpecification),
		ltSpecTimeAware, luserToNetworkRequirements,
		linterfaceCapabilities); }
\end{lstlisting}
\caption{Parsing function for SRP talker group.} \label{fig:spicy-srp}
\vspace*{-0.5cm}
\end{figure}

% example \textit{spicy} function
Fig.~\ref{fig:spicy-srp} shows an example \textit{spicy} function to parse the talker information from an SRP frame. It extracts various traffic specifications and requirements given by a talker. Moreover, we define several FRER and SRP-specific header types, e.g., FRER R-TAG for sequence numbers, to be used within the parsing functions. In the source code, the files with \textit{spicy} extension define the header types and parsing functions. 

\subsubsection{Broker} 
Broker is the built-in publish/subscribe messaging framework of Zeek. We implement three event topics in the broker: FRER, SRP talker, and SRP listener. Those topics are defined in \textit{TSN.evt} in the source code. Whenever a respective type of frame is received, the broker publishes its content, which is provided by the parser. The notice and detection engines subscribe to those topics and obtain the content of the frames for further analysis accordingly. 
% zeek broker
For \texttt{TSNZeek}, we used Zeek Broker v2.2.0.

\subsubsection{Notice Engine} \label{sec:notices}

The notice engine flags certain security events and logs received TSN traffic. We used the built-in notice facility of Zeek for this module. 
% logging event
We configure the notice engine to log the received FRER and SRP frames partially to avoid an excessive amount of logs.
% notices
A security event could be an anomaly in the configuration and network behavior, or a detected attack. The detection engine recognizes those events and then respective notice alerts are raised by the notice engine. The available notices are listed as follows. 

\begin{itemize}[wide = 0pt]
\item \textbf{N1.SRP. Excessive resource request:} If any talker demands more network resources than a predefined threshold, the notice engine raises this notice. 
\item \textbf{N2.SRP. Deviating resource request:} This notice alerts the resource demands that are marginally different from the previous SRP requests as it may indicate a malicious reservation.
\item \textbf{N3.SRP. Too many requests:} It alerts if too many requests are received in a time interval, as it can indicate a stuck talker or an attack to exhaust the network resources. 
\item \textbf{N4.SRP. Changing existing allocation:} This notice alerts in case of an attempt to change an existing SRP reservation.
\item \textbf{N5.SRP. Dangling resources:} This notice alerts the dangling resources if they are still not used after a predefined time after their registration, as it may indicate a faulty endpoint. 
\item \textbf{N6.FRER. Out of order frames:} An out of order FRER frame might indicate a malicious packet injection or a faulty endpoint. This notice alerts any out of order frame and if it should be dropped following the same mechanism as the stream recovery function in the corresponding TSN bridges.
\item \textbf{N7.FRER. Excessive member streams:} FRER duplicates member streams based on the configured degree of redundancy. This notice is raised if the number of received duplicate packets for a stream is more than the degree of redundancy. 
\item \textbf{N8.FRER. Terminated member streams:} This notice alerts if a member stream is not active, as it may indicate a node or link failure as well as an attack.
\end{itemize}

\begin{table*}[ht!]
\vspace*{0.05in}
    \caption{The overview of the notices, attack detection functions, and other aspects of the intrusion detection component.}
    \label{tab:design_recap} 
    \centering
    \begin{tabular}{l|l|l|c|c|c|c|c|c|c} \hline
    \multirow{2}{*}{\textbf{Protocol}}  &   \multirow{2}{*}{\textbf{Detection}} &   \multirow{2}{*}{\textbf{Notices}}& \multicolumn{2}{c|}{\textbf{Frequency}} &  \multicolumn{3}{c|}{\textbf{Deployment}} &  \multicolumn{2}{c}{\textbf{Context}}  \\ \cline{4-10}
    & & & \textbf{Per-frame} & \textbf{Period.}  & \textbf{Central.}  & \textbf{Local}  & \textbf{Edge}  & \textbf{Manual}  & \textbf{Stateful}  \\ \hline
    \multirow{4}{*}{\textbf{SRP}} &  A1.SRP  & N1.SRP, N2.SRP & \checkmark & \textendash & \checkmark & \textendash & \checkmark & Resource threshold & Reservations \\ \cline{2-3}
    & A2.SRP & N1.SRP, N2.SRP, N3.SRP & \checkmark & \textendash & \checkmark & \textendash & \checkmark & Rate-limit & \textendash \\\cline{2-3}
    & A3.SRP & N1.SRP, N2.SRP, N4.SRP & \checkmark & \textendash & \checkmark & \textendash & \checkmark & \textendash & Reservations \\\cline{2-3}
    & A4.SRP & N5.SRP & & \checkmark & \checkmark & \checkmark & \checkmark & \textendash & Reservations \\\hline         
            
    \multirow{3}{*}{\textbf{FRER}} & A5.FRER & N6.FRER, N7.FRER & \checkmark & \textendash & \textendash & \checkmark & \textendash & \textendash & Seq. numbers \\ \cline{2-3}
    &  A6.FRER & N7.FRER & \checkmark &\textendash & \textendash & \checkmark & \checkmark & \textendash & Seq. numbers \\\cline{2-3}
    &  A7.FRER & N7.FRER, N8.FRER & \checkmark & \checkmark & \textendash & \textendash & \checkmark & Timeout & Seq. numbers \\\hline
    \end{tabular}
\vspace*{-0.32cm}
\end{table*}

\subsection{Intrusion Detection Component} \label{sec:idc}

The intrusion detection component consists of the detection engine and another broker to communicate with the monitoring module. We implemented the detection engine purely in Python v3.9.2. 
% detection engine
It performs traffic analysis to (i) keep the current states of different streams and their configurations, (ii) make per-frame or periodical examinations to detect potential anomalies. Accordingly, it publishes the respective alerts via the broker to be logged by the notice engine. 
% broker
Note that while the first broker (attached to the monitoring component in Fig.~\ref{fig:architecture}) disseminates the frames from the event engine to others, this one establishes the communication between the notice and detection engines. It enables us to design the intrusion detection component as a standalone module that can be replaced by any other intrusion detection logic. 

The detection engine in this component introduces a set of functions to detect various SRP and FRER threats listed in~\cite{Ergenc2021b}. These functions are analogous to the rules in a rule-based IDS.
% extendable
Therefore, they are extendable to detect further threats simply as adding new rules.
In the remaining of this section, we describe the detection functions together with the attacks they can recognize. We also note on the alternative placements of \texttt{TSNZeek} in the network, i.e., centralized, local, or peripheral in the network to detect the described attacks.

\begin{itemize}[wide = 0pt]
\item \textbf{A1.SRP. Unusual SRP request:} An attacker can send malicious SRP requests to a CNC or an edge TSN bridge such as (a) demanding a bulky network bandwidth for a stream or (b) registering several streams to exhaust available resources. The detection engine detects such scenarios by comparing the requested stream traffic specifications extracted from the \texttt{TrafficSpecification} header of SRP frames with predefined threshold values for the maximum bandwidth and frame rates. It also keeps the rolling average of those values to recognize if an attacker requests stream reservations whose traffic characteristics significantly differ from the average. 
\item \textbf{A2.SRP. Flooding SRP requests:} An attacker can flood SRP requests to exhaust available resources quickly. The detection engine limits the rate of incoming requests and alerts for excessive requests. The rate limit is predefined and configured by the administrator. 
\item \textbf{A3.SRP. Changing existing allocation:} An attacker can forge an SRP request for an already registered stream to (i) reduce its reserved resources to degrade its service quality or (ii) increase its reserved resources to exhaust available resources without injecting any new stream that could be easily recognizable otherwise. The detection engine can automatically deduce if a request is accepted by checking the \texttt{TalkerStatus} group header of an SRP response. Then, for each SRP request, it checks if there already exists an accepted stream and alerts for one of the scenarios above. 
\item \textbf{A4.SRP. Dangling resources:} An attacker can reserve network resources to manipulate resource utilization without sending any real data traffic since it can also be detected and filtered by firewalls or network policies. For such cases, the detection engine periodically checks if the reserved resources are in-use. It alerts for the streams without any processed frames within a predefined time threshold. 
\end{itemize}
% deployment
\texttt{TSNZeek} should monitor all SRP traffic to detect the listed SRP-specific attacks. Therefore, while a centralized SRP configuration~(see Section~\ref{sec:srp}) imposes a centralized \texttt{TSNZeek} deployment, a distributed one requires monitoring edge bridges. 

\begin{itemize}[wide = 0pt]
\item \textbf{A5.FRER. Forging fake sequence numbers:} If an attacker can observe the current sequence number of a FRER stream, it can inject malicious frames with that sequence number so that the legit frame would be dropped by the sequence recovery function in TSN bridges. 
% for the first case
If the attacker injects a frame with the upcoming sequence number, the detection engine alerts when it detects more than one frame with the same sequence number. Besides, it deduces the expected degree of redundancy, i.e., the number of expected duplicate frames for a stream, by processing the \texttt{NumSeamlessTrees} header in the \texttt{UserToNetworkRequirements} group header of the SRP request during registration of the respective stream. 
% for the second case
If the attacker searches for the legit sequence number by sending frames with the random sequence numbers, the detection engine raises an alert for an out of order frame. 

% implementation
The detection engine mimics the stream recovery function of FRER to keep track of the legitimate intervals of the expected sequence numbers. Thus, it needs to be configured with the same recovery function, match or vector recovery used by the TSN bridges in the system. This also requires monitoring TSN bridges locally to detect where exactly malicious frames are injected and dropped.
\item \textbf{A6.FRER. Malicious rerouting:} If there are intersections between redundant paths, it eliminates duplicate packets being forwarded through the same bridge~\cite{Ergenc2021a}. Instead of directly sabotaging the communication, an attacker could subtly reroute the redundant stream through intersecting paths to force FRER to drop the duplicate packets and hinder the redundancy. Therefore, \texttt{TSNZeek} examines the configured FRER routes, e.g., against malicious misroutings, intersecting paths etc. This requires the \texttt{TSNZeek} attached to the SRP controller, i.e., centralized or hybrid, to access to the configuration of redundant paths.
\item \textbf{A7.FRER. Triggering timeout:} An attacker can enforce FRER functions on TSN bridges to raise a \texttt{RECOVERY\_TIMEOUT} event~\cite{ieee-802.1cb} if it can block all member streams of a FRER stream. Consequently, the expected sequence number of that stream is revoked. Once the attacker sends the first frame after this event, it becomes the valid originator of the stream with the forged initial sequence number. The detection engine recognizes the absence of the original member streams by measuring the time passed after the reception of the last frame of the respective streams. It also detects if the same stream has a new sequence number by per-frame examinations. Both can be detected by monitoring the edge bridge that the destination endpoint is attached to.
\end{itemize}

Table~\ref{tab:design_recap} summarizes attack detection and notices with several related aspects discussed above. These aspects are (i) how frequently \texttt{TSNZeek} investigates an attack, i.e., per-frame or periodically, (ii) how \texttt{TSNZeek} instances should be deployed to detect an attack, i.e., centralized, local, or to the edge, and (iii) what \texttt{TSNZeek} needs for the detection, i.e., manual configuration and the current state of stream reservations.

\section{Evaluation} \label{sec:experiments}

This section presents our experimental setup and evaluation results for the efficiency and performance of \texttt{TSNZeek}.

\subsection{Experimental Setup} \label{sec:setup}

\begin{figure}[t!]
  \centering
  \includegraphics[width=0.4\textwidth]{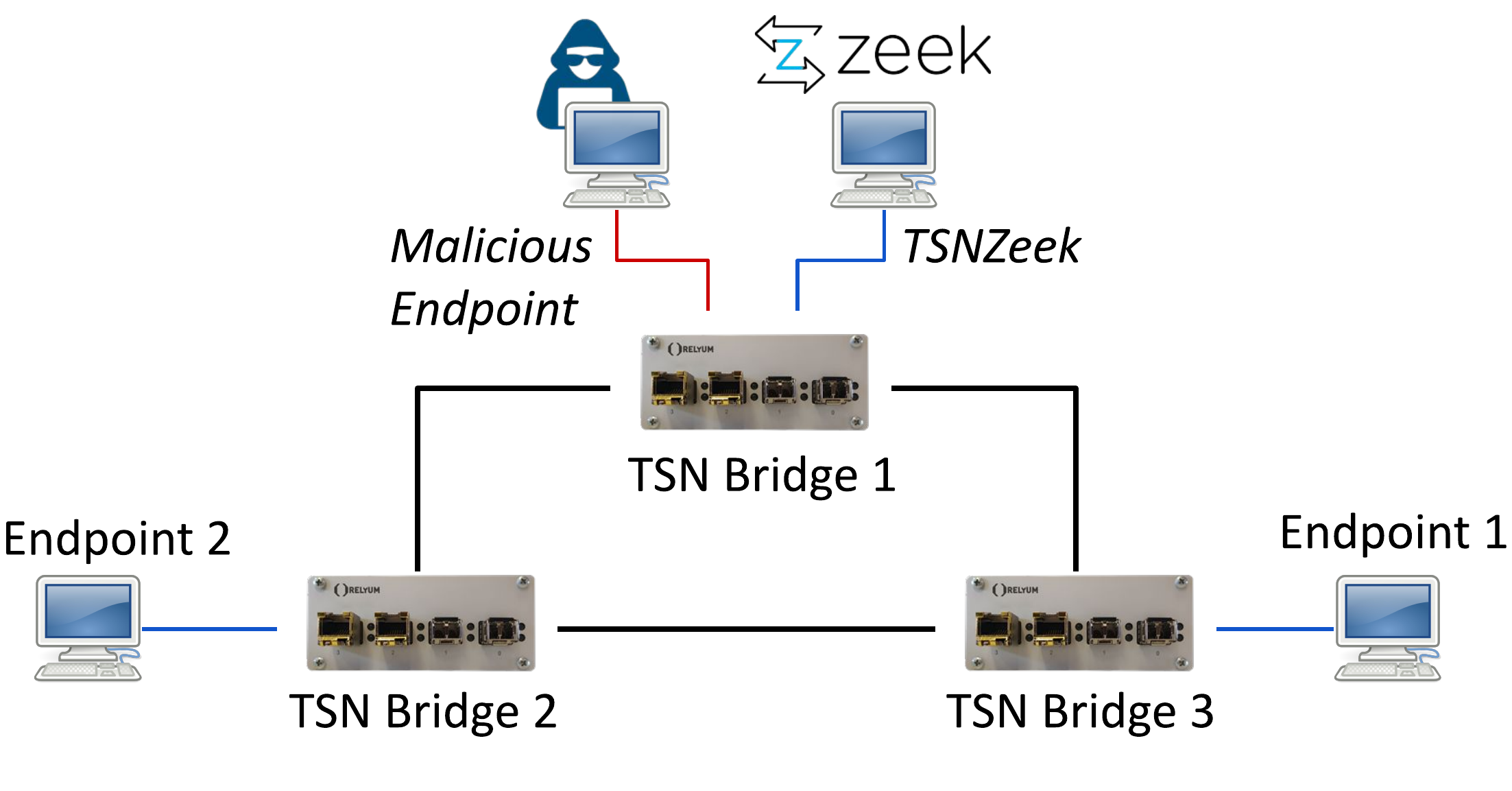}
    \caption{Testbed setup. \vspace*{-0.20in}}
  \label{fig:topology}
\end{figure}

For our experiments, we set up a real TSN testbed shown in Fig.~\ref{fig:topology}.
% describing the setup
It consists of three TSN bridges~(TSN1, TSN2, and TSN3) connected in a ring topology. An endpoint (EP1) is attached to TSN3, and another (EP2) is attached to TSN2. \texttt{TSNZeek} is deployed on a computer with an Intel Core i3-9100 3.60Ghz CPU and connected to TSN1. A malicious endpoint~(MEP) is also attached to TSN1 to conduct attacks.  
% describing the test scenario 
In our application scenario, EP1 first sends an SRP request for a resource reservation to stream a video. Then, it sends the data over two redundant paths, TSN2-TSN1-TSN3 and TSN2-TSN3, to EP2 and EP3 using FRER.
% monitoring
\texttt{TSNZeek} monitors FRER frames forwarded over TSN1. 

Since there is not any public TSN dataset including malicious traffic, we also implemented attacks described in Section~\ref{sec:idc} in Python (available in the source code). 

\subsection{Results}

We evaluated the resource usage and intrusion detection capabilities of \texttt{TSNZeek}. For resource usage, we measured the CPU utilization of the monitoring and detection components, as well as the packet processing rate and delay of \texttt{TSNZeek}. For intrusion detection, the typical evaluation metrics for an IDS, e.g., accuracy, sensitivity~\cite{Tidjon2019}, are not directly fitting for our rule-based IDS as its objective is detecting the specific threats according to the implemented rules. Therefore, we tested the effectiveness of the detection module against the attacks described in Section~\ref{sec:idc}. 

\begin{figure}[b!]
  \vspace*{-0.15in}
  \centering
         \includegraphics[scale=0.35]{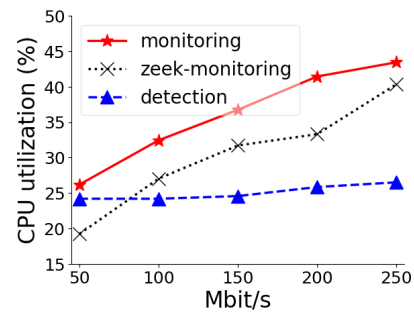}
         \caption{CPU utilization of the monitoring and detection components.}
  \label{fig:cpu}
  \vspace*{-0.15in}
\end{figure}

\subsubsection{Resource usage} 

Since IEEE 802.1 TSN protocols define data link layer protocols, processing the events starting from such low-level communication may easily lead to high resource usage. Accordingly, we measured the CPU usage of \texttt{TSNZeek} for an increasing data load from 50 to 250 Mbit/s. This interval of data load is reasonable for the number of critical streams in a TSN network and mainly limited by the processing capacity of the TSN bridges in our testbed. We generated the load using \textit{iperf}, which is a network speed test tool. Fig.~\ref{fig:cpu} shows the resource usage of the monitoring component (Zeek module extended with TSN grammar, red and solid line), the detection component (Python program running at the control plane, blue and dashed line), and also the CPU consumption of native Zeek without TSN support (processing non-TSN Ethernet frames, black and dotted line). When increasing the traffic load, the CPU utilization of the monitoring component increases from 25\% at 50Mbit/s to 45\% at 250Mbit/s. As shown in the figure, \texttt{TSNZeek} consumes only $\sim$5\% more CPU power than the Zeek instance without TSN support. The detection module has a constant resource utilization of around 25\% as it only processes singular events that are sent by the monitoring module.  

\begin{figure}[t!]
  \centering
         \includegraphics[scale=0.33]{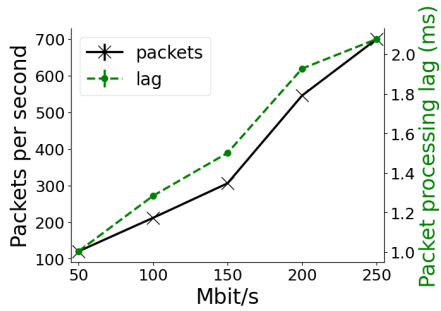}
         \caption{Packet processing performance of the monitoring component.\vspace*{-0.2in}}
  \label{fig:packet}
\end{figure} 

Fig.~\ref{fig:packet} shows the packet processing rate and lag of \texttt{TSNZeek}. The packet processing lag describes the time passed between the reception and parsing of a frame. The figure shows that the packet processing rate~(black, solid line) increases proportionally with an increasing data load without any packet drops, e.g., due to a potential congestion. Increasing load also leads to a higher packet processing lag of up to 2 ms (green, dashed line). For any lag in milliseconds, time-sensitive frames with submillisecond latency requirements may already be delivered before an intrusion alert. Although it is not critical for a monitoring module, a potential time-sensitive intrusion \textit{prevention} system utilizing \texttt{TSNZeek} might require further improvements in data processing speed. 

\begin{figure}[ht!]
  \centering
\begin{lstlisting}[language=json,  firstline=2, lastline=7, basicstyle=\footnotesize]
    "date": "2022-10-26-16-41-02",
    "timestamp": 1666802462.036670,
    "note": "TSN::POTENTIAL_ATTACK_6",
    "protocol": "FRER",
    "msg": "Out of order frame is discarded for stream 29695 - seq.nr. 7148 < 54972",
    "actions": "Notice::ACTION_LOG"
\end{lstlisting}
    \caption{A sample intrusion alert in \textit{json} format.}
  \label{fig:sample-log}
  \vspace*{-0.1in} 
\end{figure} 

\subsubsection{Intrusion detection} 

In our experiments, \texttt{TSNZeek} can successfully detect all the listed attacks in Section~\ref{sec:idc} and raise the respective notices in real-time. Fig.~\ref{fig:sample-log} shows N6.FRER (in json format) in the log stream of the notice engine against the attack A5.FRER, i.e., the frame injection with the sequence number 7148, while the expected one is 54972. 

However, we still observe redundant notifications in particular scenarios.
% out of order
For instance, when we connect an Ethernet hub between TSN1 and TSN3, e.g., extending the network with a non-TSN network component, a member stream delivers out of order packets due to the delayed frames. \texttt{TSNZeek} notices this as a malicious attempt because of highly deviating sequence numbers. Although this is unusual for strictly configured TSN systems, \texttt{TSNZeek} should still be configured considering such network conditions.

\section{Conclusion} \label{sec:conclusion}
Although IEEE 802.1 TSN standards propose emerging time-sensitive communication protocols for critical systems, they still lack security countermeasures against potential attack vectors. In this paper, we present the first open-source security monitoring and intrusion detection system, \texttt{TSNZeek}, for IEEE 802.1 TSN protocols. We implement \texttt{TSNZeek} by extending an existing monitoring tool, Zeek, with a new packet parsing grammar and the stream analysis engines addressing TSN-specific security events and attacks.  
We evaluate its resource usage and confirm that it can successfully detect various attacks against the prominent TSN protocols, SRP and FRER. For future work, we aim to improve our detection engine to minimize redundant and false alerts by accurately modeling the usual TSN behavior.

\vspace*{-0.02in}
%\balance
\bibliographystyle{ieeetr}
\bibliography{references}
\end{document}